\documentstyle[11pt,epsfig]{article}

\begin{document}

\begin{titlepage}

\begin{flushright}
CERN-PH-TH/2004-172\\
\end{flushright}

\vspace{1cm}

\begin{center}

\huge{Old ideas and new twists \\ in string cosmology}

\vspace{1cm}

\large{Massimo Giovannini}

\normalsize
\vspace{.1in}

{\sl Department of Physics, Theory Division,\\
 CERN, 1211 Geneva 23, Switzerland }

\vspace*{1cm}

\begin{abstract}
Some of the  phenomenological implications of string cosmological models 
are  reviewed, with particular attention to the 
spectra of the tensor, scalar and vector modes of the geometry.
A class of self-dual string cosmological models is presented. 
These solutions provide  an effective description of cold bounces, where 
a phase of accelerated contraction smoothly evolves into an epoch of  
decelerated Friedmann--Robertson--Walker expansion dominated by the dilaton. 
Some of the general problems of the scenario (continuity of the perturbations, 
reheating, dilaton stabilization,...) can be 
successfully discussed in this framework.
\end{abstract}
\vspace*{2.5cm}
{\em Invited review  
to appear in the Proceedings of the\\ ``Third International 
Conference on Frontier Science'', \\
INFN National Laboratories, Frascati (Rome), June 14--19, 2004}
\end{center}
\end{titlepage}

\newpage

\parskip 0.2cm

\section{Introduction}
Heeding observations, the large-scale temperature fluctuations 
detected in the microwave sky are compatible with a 
quasi-flat spectrum of curvature inhomogeneities. Quasi-flat means, in the 
present context, 
that the Fourier transform of the two-point function
of the scalar fluctuations of the geometry
depends on the comoving momentum $k$ as  $ k^{n_{s} -1}$, with 
$ n_{s} \simeq 1$. Taking only the WMAP determination of $n_{s}$
\cite{WMAP1,WMAP2}, the scalar spectral index lies in a rather narrow range,
$0.95 \leq n_{s} \leq 1.03$.
For later convenience, if $n_{s} \geq 1$ the spectra are said to be 
 blue, if 
$n_{s} < 1$ the spectra are said to be  red and, finally, 
 if $n_{s} \gg 1$ the power spectra are  called violet.

From the observations at  smaller angular scales, 
it is by now established that  the temperature fluctuations exhibit  
oscillations (the so-called Sakharov or Doppler oscillations) as a function 
of the sound horizon at decoupling. From the 
typical structure of these oscillations it is possible to argue that 
the curvature fluctuations present outside the horizon after 
equality (but before decoupling) were also constant.
If the  curvature fluctuations are constant, the solutions of the  
evolution of the density 
contrasts and of the peculiar velocities for the various species present 
in the plasma imply that the fluctuations in the total  entropy density 
vanish at large-distance scales. These initial conditions for the evolution of the CMB 
anisotropies are often named {\em adiabatic}.  

If a single field drives the conventional inflationary dynamics, 
 the scalar fluctuations of the geometry naturally have  a quasi-flat 
spectrum and are also constant at large-distance scales after 
matter--radiation equality. 
The quasi-flatness of the spectrum is related, in these models, to the quasi-constancy 
of the Hubble expansion rate and of the Ricci scalar during the inflationary 
stage. More precisely, in the 
context of single-field inflationary models, the  
 curvature scale has a monotonic behaviour as a function of
(cosmic or conformal) time coordinate, and it  
is always (slowly) decreasing. Since the curvature scale decreases, 
it can be argued, on a rigorous basis, 
 that a true physical singularity is present 
in the far past \cite{alex}. However, the 
dynamics of the initial singularity is screened by the 
long period of inflation,  during which 
the possible gradients arising in the matter fields are diluted and eventually 
erased if the duration of inflation exceeds 65-efolds 
(see for instance \cite{brand}). 

In the context of string cosmological models the conventional 
inflationary scenario seems 
quite difficult to obtain and therefore, during the last fifteen years, 
various cosmological models inspired by string theory have been explored.
One of the features of these models is that the curvature scale 
is far from being constant but 
it is rather steeply increasing, at least during a sizeable portion 
of the early history of the Universe. 
In these models {\em a singularity}  
is often encountered 
just after the phase of growing curvature and gauge coupling. 
This problem is not an easy one 
to address be it technically or conceptually. 
Owing to the mentioned phase of growing curvature, the perturbation spectra 
obtained in string cosmological 
models are far from being quasi-flat. 
They are, indeed, rather violet.

There are by now several variations on
this pre-big  bang theme. Besides the original pre-big bang (PBB)
scenario \cite{GV1,GV2}, based on the duality symmetries of string
cosmology,  new models incorporating brane and M-theory ideas
have been proposed under the generic name of ekpyrotic (EKP)  scenarios
\cite{EKP1,EKP2}.
The various  proposals differ in the way the scale factor behaves 
during the growing-curvature phase. However, they  all share the  feature
of describing a bounce in the  space-time
curvature. A common theoretical challenge to all these models is
that of being able to describe the transition between the two regimes.

\section{Tensor, scalar and vector modes\\ in string cosmological models}

String cosmological models are naturally
 formulated in more than four dimensions. This 
occurrence implies that the fluctuations of a higher dimensional geometry may 
be more complicated than a simple four-dimensional space-time. 
However, in order to simplify the 
discussion, let us consider, as  was  done in the past, the dimensionally 
reduced low-energy string effective action, which can be written as 
\begin{eqnarray}
&& S_{\rm eff} - \int d^{4} x \sqrt{- G} e^{- \varphi}\biggl[ 
\frac{1}{2 \lambda_{s}^2}\biggl( R + G^{\alpha\beta} \partial_{\alpha} \varphi 
\partial_{\beta} \varphi + V(\varphi)- 
\frac{1}{12} H_{\mu\nu\alpha} H^{\mu\nu\alpha} \biggr) 
\nonumber\\
&& + \frac{1}{4} F_{\mu\nu} F^{\mu\nu} + ...\biggr].
\label{action}
\end{eqnarray}
which is the typical outcome of the compactification of 
ten-dimensional superstrings on a six-dimensional torus. 
A few specifications  are in order about this expression
\begin{itemize}
\item{} $\lambda_{s}$ is the string length scale,  related to the  
Planck length scale  by  $\ell_{\rm P} = e^{\varphi/2} \lambda_{s}$;
\item{} $\varphi = \Phi_{10} - \ln V_{6}$ is the four-dimensional 
dilaton field, which can be expressed in terms of the ten-dimensional 
dilaton $\Phi_{10}$ and in terms of  the volume of the six-dimensional torus;
\item{} $V(\varphi)$ is the four-dimensional dilaton potential;
\item{} $H^{\mu\nu\alpha}$ is the antisymmetric tensor field,
 related, in four dimensions, 
to a pseudo-scalar field $\sigma$ by $H^{\mu\nu\alpha} = e^{\varphi}
 \epsilon^{\mu\nu\alpha\rho}/\sqrt{-G} \partial_{\rho}\sigma$;
\item{} $F_{\mu\nu}$ is a generic Abelian gauge field;
\item{} the ellipses stand for other fields 
(other gauge fields, both Abelian and non-Abelian, 
chiral fermions,...) and for the corrections, which can be both 
of higher order in $\lambda_{s}^2 \partial^2$ (higher derivatives 
expansion producing  quadratic corrections to the Einstein--Hilbert 
action) and of higher order in $e^{\varphi}$ (loop expansion).
\end{itemize}
Equation (\ref{action}) is written in the so-called string frame metric 
where the Ricci scalar $R$ is coupled to the 
four-dimensional dilaton. Other frames can be employed by
appropriately   redefining the metric and the dilaton. 
A particularly useful frame is the Einstein frame, where the Ricci 
scalar is not directly coupled to $\varphi$.

\begin{figure}  
\begin{center}
\epsfig{figure=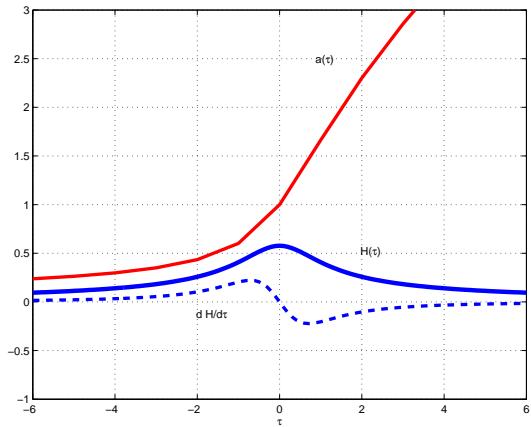,width=7cm}  
\end{center}
\caption{The evolution of the Hubble expansion rate  $H$, of its derivative, 
and of the scale factor $a(\tau)$ 
in the cold bounce models. The variable $\tau = t/t_{0}$ is  
the cosmic time coordinate (in the string frame)
rescaled by the typical time scale of the bounce, $t_0$. } 
\label{fig1}
\end{figure}

As already mentioned, the evolution equations for the metric 
and the dilaton can lead to singular 
solutions. This situation is, however, not generic, since also non-singular 
solutions can be  found \cite{maxbounce1,maxbounce2,maxheating}
and an example  is reported  Fig. \ref{fig1}. The solutions 
illustrated in there can be derived in the presence 
of a dilaton potential, that depends directly, not on $\varphi$, but 
on $\overline{\varphi} = \varphi - \log{\sqrt{-G}}$, i.e. the shifted dilaton 
usually defined in the context of the $O(d,d)$-covariant description 
of the low-energy string effective action \cite{meis}. 
From the point of view of the Einstein frame dynamics, 
these solutions describe a phase of accelerated 
contraction, evolving smoothly into an epoch of decelerated expansion 
\cite{maxheating}. 
The two regimes of the solution are connected, 
by scale-factor duality \cite{GV1}. 

\subsection{Tensors}
\begin{figure}  
\begin{center}
\epsfig{figure=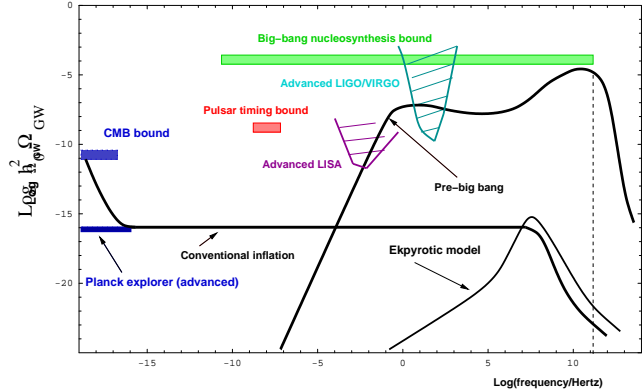,width=8.5cm}  
\end{center}
\caption{The spectrum of relic gravitons from various 
cosmological models expressed in terms 
of $\Omega_{\rm GW}$ ($h_{0}$ is the experimental indetermination 
on the present value of the Hubble expansion rate). } 
\label{fig2}
\end{figure}

The spectrum of the tensor modes arising from solutions  where the Hubble expansion rate 
is increasing 
has been computed in various steps \cite{pbbgw1}. 
 The amplified tensor modes of the geometry 
lead to a stochastic background of gravitational waves (GW)
with  violet spectrum both 
in the GW  amplitude and energy density. This expectation is confirmed 
also in the context 
of the models illustrated in Fig. \ref{fig1} as well as  in the context 
of other 
non-singular models. In Fig. \ref{fig2} the  GW signal is parametrized 
in terms of the logarithm of
$\Omega_{\rm GW} = \rho_{\rm GW}/\rho_{\rm c}$, i.e. the fraction of 
critical energy density 
present (today) in GW. On the horizontal axis of Fig. \ref{fig1} the 
logarithm of the present (physical) 
frequency $\nu$ is reported.
In conventional inflationary models, for $\nu \geq 10^{-16} $ Hz, 
 $\Omega_{\rm GW}$, is constant (or slightly decreasing)  
as a function of the present 
frequency. In the case of string cosmological models $\Omega_{\rm GW}
 \propto \nu^3 \ln{\nu}$, which also 
implies a steeply increasing power spectrum.
This possibility spurred various experimental groups to analyse possible directs 
limits on the scenario arising from specific instruments such as  resonant 
mass detectors \cite{bars} 
and microwave cavities  \cite{picasso,cruise}. These attempts are justified since the
signal of pre-big bang models may be rather strong at high frequencies and, anyway, much stronger 
than the conventional inflationary prediction
 
The sensitivity of a pair of VIRGO detectors
to string cosmological gravitons has been  specifically analysed
 \cite{maxdan} with the 
conclusion that a VIRGO pair, in its upgraded stage, can certainly  probe 
wide regions of the parameter space of these models. If we  maximize the 
overlap between the two detectors \cite{maxdan} or 
if we reduce (selectively) the pendulum and pendulum's internal modes
contribution to the thermal noise of the instruments, the 
visible region (after one year of observation and with ${\rm SNR} = 1$)
of the parameter space will get even larger. Unfortunately, as in the 
case of the advanced LIGO detectors, the sensitivity to a flat $\Omega_{\rm GW}$ will be irrelevant for 
ordinary inflationary models also with the advanced VIRGO 
detector. It is worth mentioning 
that growing energy spectra of relic gravitons  can also arise 
in the context of quintessential inflationary models \cite{alex2,maxquint}.
In this case $\Omega_{\rm GW} \propto \nu \ln{\nu}$ (see \cite{maxquint} 
for a full discussion). 

In order to gauge carefully our theoretical expectations it is relevant to notice that 
direct experimental limits on stochastic GW backgrounds are rather 
far from the interesting region of the parameter space of a possible cosmological signal.
In particular, form various instruments (resonant mass detectors, interferometers) 
$h_{0}^2 \Omega_{\rm GW} < {\cal O}(10)$. From Fig. \ref{fig2} it can be easily 
appreciated that a cosmological signal must satisfy $h_{0}^2 \Omega_{\rm GW} < {\cal O} (10^{-4})$
as implied by the bound on extra-relativistic species at big-bang nucleosynthsis. This 
constraint is, however, model-dependent and it can be slightly relaxed in unconventional 
models of big-bang nucleosyntheis \cite{maxhannu} where more extra-relativistic species 
are allowed since the presence of matter--antimatter domains allows an independent reduction of the 
$^{4} {\rm He}$ abundance (which is led to increase by the presence of extra-relativistic species). 
 \begin{figure}  
\begin{center}
\epsfig{figure=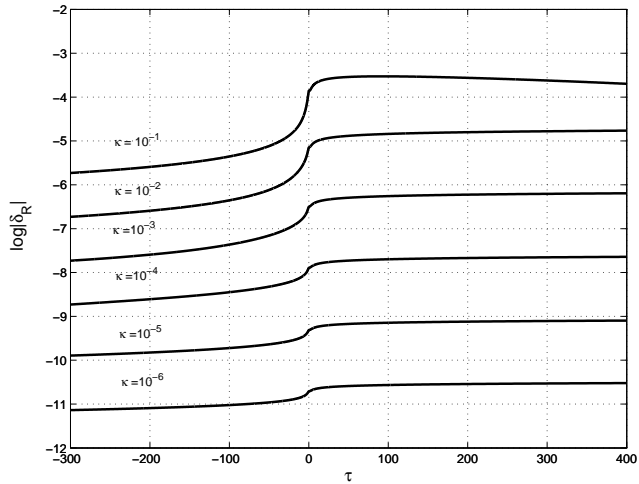,width=8.5cm}  
\end{center}
\caption{ The Fourier transform of the two-point function of ${\cal R}$  
computed for different comoving momenta as a function of the cosmic time 
coordinate. } 
\label{fig3}
\end{figure}
 
\subsection{Scalars}
The spectrum of the tensor modes of the geometry is not controversial   
because the
tensor fluctuations of the geometry are defined as a rank-2 (divergenceless 
and traceless)  tensor in 
three dimensions. Consequently they  are invariant under infinitesimal 
coordinate transformations. 
Scalar perturbations, in contrast, do depend on the specific coordinate 
system and are described, in four dimensions, by a single 
propagating degree of freedom. This 
problem is only partially alleviated by the possibility of defining variables 
that are invariant under coordinate transformations. In fact, 
different choices are equally allowed such as:
\begin{itemize}
\item{} Bardeen potentials (curvature perturbations on shear 
free hypersurfaces), usually denoted by $\Psi$,
\item{} curvature perturbations on comoving hypersurfaces, usually 
denoted by ${\cal R}$,
\item{} curvature perturbations on constant-density hypersurfaces, 
usually denoted by $\zeta$,
\end{itemize}
and as well as other choices. 
The  variables listed above are related by specific 
differential relations: once one of them is reliably 
computed, all the other follow. 

Using various descriptions (both gauge-invariant and gauge-dependent), 
 it was argued in \cite{pbbscal1}
that the spectra of scalar fluctuations are also violet, with a 
scalar spectral index $n_{s} = 4$.
The same analysis, applied to the case of ekpyrotic models, would lead 
to $n_{s}= 3$.
The pre-big bang estimates \cite{pbbscal1} have been questioned on various 
grounds. The bottom line of these arguments 
would be that, in single field pre-big bang or ekpyrotic models, the spectrum 
of the tensor modes is violet but 
the spectrum of the scalar modes may be flat or even red, i.e. increasing at 
large-distance scales. 
The analysis of the models illustrated in Fig. \ref{fig1} seems to give 
an unambiguous answer: while the evolution of the Bardeen potential is rather complicated 
around the bounce the time dependence  of both ${\cal R}$ and $\zeta$ is rather smooth.
 Furthermore, not only the spectrum 
of ${\cal R}$ and $\zeta$ is, as expected, violet, but it is also in agreement with the analytical estimate.
The results for the evolution of ${\cal R}$ is illustrated in Fig. \ref{fig3}. The value of the 
comoving momentum increases from bottom to top. Hence, the spectrum is increasing 
as a function of $k$, as expected. An accurate numerical determination discussed in \cite{maxbounce2},
also shows that $\delta_{{\cal R}} \sim k^{3/2}$ with specific logarithmic corrections.  The spectrum 
of the Bardeen potential has  also been computed accurately in \cite{maxbounce2}, with the 
result that $\delta_{\Psi} \sim k^{-1/2}$ as expected from the analytical estimates.

\subsection{Vectors}

Vector modes 
of the metric are not excited in the context of conventional inflationary models. If the background 
geometry has more than four dimensions,  on the other hand vector modes are  expected \cite{maxint}. 
It is also possible to envisage the situation where  rotational modes of the geometry are excited 
by the fluctuations of the velocity field \cite{brandvec}. The cold-bounce solutions illustrated 
in Fig. \ref{fig1} can be generalized to include fluid sources \cite{maxheating} as well as internal
(contracting) dimensions \cite{maxvector}.
 It was recently argued that vector modes of the geometry can be 
produced in pre-big bang and ekpyrotic/cyclic scenarios \cite{brandvec}. In particular, it was argued 
that the vector modes of the geometry may lead to a growing mode prior to the 
occurrence of the bounce. This expectation has been verified in \cite{maxvector} but it has also 
been shown that, in  four dimensions, the growing vector mode present before the bounce turns into a decaying mode after the bounce.
This result has been achieved in specific models of smooth evolution similar to the ones 
presented in Fig. \ref{fig1} but including also fluid sources. It will be interesting 
to analyse more precisely this problem in different both singular and non-singular models. 
Going  beyond four dimensions, the vector modes of the geometry are copiously produced \cite{maxvector}. The 
higher-dimensional examples provided in \cite{maxvector} support the evidence that, in  semi-realistic models these 
spectra may be red.
\subsection{Heating up the cold bounce}
The cold-bounce solutions discussed in \cite{maxbounce1} and \cite{maxbounce2} and illustrated in Fig. \ref{fig1} 
 certainly have realistic 
features. However there are two less realistic aspects
\begin{itemize}
\item{} after the bounce the Universe is cold and dominated by the dilaton field;
\item{} the dilaton field is not stabilized in the sense that it does not reach constant 
value.
\end{itemize}
These two problems may be solved if the back-reaction of the various massless fields (i.e. massless 
gauge bosons, for instance) is included. In \cite{maxheating} the back-reaction effects of 
high-frequency photons has been included and the following results have been illustrated:
\begin{itemize}
\item{} an accurate numerical method for the calculation of the 
amplification of the primordial photons has been developed;
\item{} taking into account the back-reaction of the primordial photons,
the cold-bounce solution can be consistently heated;
\item{} the transition to radiation can occurr for sub-Planckian 
curvature scales;
\item{} solutions have been presented where the dilaton goes to a constant 
value and the asymptotic value of the gauge coupling is always smaller than 
$1$.
\end{itemize}

\section{A consistent phenomenological framework}

In the PBB case it was admitted early on  that the tensor \cite{pbbgw1} and 
adiabatic- 
curvature perturbations \cite{pbbscal1} had too large a spectral index to be of any
relevance at cosmologically interesting scales (while being possibly
important for gravitational-waves searches \cite{maxdan}). Isocurvature
perturbations (related to the Kalb--Ramond two-form)  can instead  be
produced with an interestingly flat spectrum \cite{cop}, but have to be
converted into adiabatic-curvature perturbations through the so-called
curvaton mechanism \cite{enq} (see also \cite{lyth,maxtracking}) before they can provide a viable
scenario for large-scale anisotropies in the pre-big bang context \cite{maxaxion1}.
Proponents of the ekpyrotic scenario, while agreeing with the PBB 
result of a steep spectrum of tensor perturbations, have also
repeatedly claimed \cite{EKP3} to obtain ``naturally" an almost
scale-invariant spectrum of adiabatic-curvature perturbations, very
much as in ordinary models of slow-roll inflation. These claims have
generated a debate (see
for instance \cite{deb}), with many arguments given  in favour of the phenomenological viability of  EKP scenarios 
 in the absence of a
curvaton's help or against it. The reasons for the disagreement  can be ultimately
traced back to the fact that the curvature bounce
 is put in by hand (see for instance \cite{copeland}), rather than  derived
from an underlying action. This leaves different authors to make
different assumptions on how to smoothly connect  perturbations across
the bounce itself, which results in completely different physical
predictions. 

A specific class of models has been illustrated \cite{maxbounce1,maxbounce2}. In these  
the evolution of the background geometry and of the dilaton coupling is regular and specific 
testable predictions are possible.
 On the basis of these semi-analytical investigations 
it can be  argued that in single-field pre-big bang models the spectrum of the scalar and tensor modes 
of the geometry is, as expected from previous estimates, violet. The situation 
 changes when the evolution of the fluctuations of the Kalb-Ramond field 
is consistently included. In this case a flat spectrum of curvature fluctuations can be obtained 
and compared with the observed anisotropies in the CMB (see \cite{maxaxion1} for a complete discussion).
The large-scale microwave anisotropy probes can then be used to constrain the various parameters 
of the model.

\section*{Acknowledgements}
It is a pleasure to thank P. Picozza and M. Ricci for their kind invitation and for 
the nice scientific atmosphere of this meeting. The author would also like to thank 
M. Gasperini and G. Veneziano for a stimulating collaboration on some of the topics 
covered by this contribution.

\end{document}